\documentclass[twocolumn,showpacs,preprintnumbers,amsmath,amssymb]{revtex4}

\usepackage{graphicx}
\usepackage{dcolumn}
\usepackage{bm}

\begin{document}

\title{Condensation-Driven Aggregation in One Dimension}%

\author{M. K. Hassan$^{\dag}$ and M. Z. Hassan$^{\$}$
}%
\date{\today}%

\affiliation{
$\dag$ University of Dhaka, Department of Physics, Theoretical Physics Group, Dhaka 1000, Bangladesh \\
$\$ $ ICT Cell, Bangladesh Atomic Energy Commission, Dhaka 1000, Bangladesh 
}

\begin{abstract}%
We propose a model for aggregation where particles are continuously growing by 
heterogeneous condensation in one dimension and solve it exactly. We show that the particle 
size spectra exhibit transition to dynamic scaling 
$c(x,t)\sim t^{-\beta}\phi(x/t^z)$. The exponents $\beta$ and $z$ satisfy a generalized scaling relation 
$\beta=(1+q)z$ where the value of $q$ is fixed by a non-trivial conservation law. We have 
shown that the value of $(1+q)$ is always less than the value
$2$ of aggregation without condensation.
\end{abstract}

\pacs{61.43.Hv, 64.60.Ht, 68.03.Fg, 82.70Dd}

\maketitle
\section{Introduction}

The formation of clusters by aggregation of particles is a characteristic feature of many 
seemingly different
processes in physics, chemistry, biology and engineering. Examples include aggregation of 
colloidal or aerosol particles suspended in liquid or gas 
\cite{ref.friedlander, ref.thorn,ref.melle}, 
polymerization \cite{ref.polymerization}, antigen-antibody aggregation \cite{ref.antigen}, 
and cluster formation in galaxy \cite{ref.galaxy}. This wide variety of applications 
has resulted in numerous  studies which reveal 
that, when chemically identical particles aggregate, almost always scale invariant clusters
emerge. Note that due to the non-equilibrium nature of the aggregation process the standard 
theoretical framework developed for equilibrium statistical physics is found redundant. 
However, the application of 
stochastic theory is found to be increasingly useful in capturing a wide class of 
non-equilibrium phenomena. 

Typically, the non-equilibrium systems are described by the rate equation approach, having
the form of a master equation which is often governed by some conservation principle. 
The Smoluchowski equation for the kinetics of irreversible 
aggregation is one such example, where the distribution function $c(x,t)$ of particle of
size $x$ at time $t$ evolves according to the following integro-differential equation 
\cite{ref.smoluchowski, ref.chandrasekhar}:
\begin{eqnarray}
\label{eq:cs}
{{\partial c(x,t)}\over{\partial t}} 
&=& -c(x,t)\int_0^\infty K(x,y)c(y,t)dy \nonumber \\ &+ & {{1}\over{2}}\int_0^x dy K(y,x-y)
c(y,t)c(x-y,t).
\end{eqnarray}
Here, the kernel $K(x,y)$ is symmetric with respect to its argument and it 
determines the collision time in which a particle of size $x$ 
collides with another particle of any size $y$ and they merge into a particle (aggregate) 
of size $(x+y)$.
The first term on the right hand side of Eq. (\ref{eq:cs}) describes the loss of a particle 
of size $x$ due to merging of particles of size $x$ with another particle of any size, while 
the second term describes gain of $x$ due to merging of
particles of size $(x-y)$ with $y$. The Smoluchowski equation has been studied extensively for 
a large class of 
kernels satisfying  $K(bx,by) = b^\lambda K(x,y)$. The homogeneity exponent $\lambda$ is shown 
to play a crucial role in classifying 
gelling and non-gelling models. For instance, $\lambda<1$ describes the non-gelling model whose dynamics
is governed by the conservation of mass principle, and $\lambda>1$ is the gelling model that
describes the gelation transition accompanied by the violation of 
mass conservation law \cite{ref.ziff,ref.lushnikov}. Note, however, that, 
despite the seemingly simple structure 
of Eq. (\ref{eq:cs}), it is solved exactly for non-gelling model, only for a constant
kernel $\lambda=0$, and that solution was given more than one hundred years ago by Smoluchowski himself. 
Finding another exact analytical solution of Eq. (\ref{eq:cs}) for the non-gelling 
model ($\lambda<1$) therefore still remains an open challenge.

One of the reason why the Smoluchowski equation was so successful is that it has provided 
much of our theoretical 
understanding about dynamic scaling and associated exponents which are in aggreement, at 
least qualitatively,
with those extracted from real experiments and numerical simulations \cite{ref.vicsek}. For 
instance, for $\lambda<1$
it was shown that the distribution function $c(x,t)$ exhibits dynamic scaling,
\begin{equation}
\label{eq:0}
c(x,t)\sim t^{-\beta}\phi\big (x/t^z\big );\hspace{0.30cm} {\rm with} \hspace{0.25cm} z>0, 
\end{equation}
in the long time ($t\rightarrow \infty$), large size ($x\rightarrow \infty$) limit where $\phi(\xi)$ 
is a scaling function whose argument $\xi=x/t^z$ is a dimensionless quantity.    
The exponents $\beta$ and $z$ satisfy a scaling relation $\beta=\theta z$ with $\theta=2$
which follows from the conservation of mass principle. A scaling form like Eq. (\ref{eq:0})
 is shared by an extraordinarily diverse range of other phenomena, not just in aggregation,
 e.g., systems exhibiting 
self-organized criticality, cluster growth in driven diffusive systems, fragmentation 
processes, etc. \cite{ref.scaleinvariance, ref.scaling}. The ubiquity of this 
scaling form suggests the existence of a common underlying mechanism  
which makes such seemingly disparate systems behave in a remarkably similar fashion.

In addition to growth by aggregation, there exists a host of other mechanisms (e.g., 
condensation, deposition, and 
accretion) whereby particles can  grow continuously between aggregations 
\cite{ref.droplet,ref.sire,ref.husar,ref.dust}. For instance, aerosol or colloidal 
particles are often not stable but evolve via aggregation and condensation, leading 
to gas-to-particle conversion. 
However, when the concentration of particles present is high and the super-saturation is 
low, the condensation is heterogeneous in nature since, in this case, condensation 
takes place only on the existing particles without forming new nuclei 
\cite{ref.friedlander,ref.husar}. Otherwise, the system may have sufficient number of 
impurities, such
as dirt or mist particles, which usually serves as potential
nucleation sites on which condensation takes place,
 and the resulting process is known as homogeneous condensation. 
In the latter case, 
the gas starts condensing on such nuclei and thereafter it is counted as a new particle, 
which then may take part on 
aggregation processes with other particle in the system. However, in the present work we 
will consider only the former case - heterogeneous condensation.  Such condensation-driven 
aggregation does play an important role in the formation of the size spectra that ultimately 
determine the various physical properties of the aggregates. Motivated by 
this, we here propose a simple model which is defined in the next section.

The rest of the paper is organized as follows. In Sec. II, we present the definition of
our model including its algorithm. Interesting results from numerical simulation
based on the algorithm are presented  in Sec. III.
In section IV, we propose a generalized version of the Smoluchowski equation
and its appropriate parameters to describe the evolution of the 
distribution function $c(x,t)$ following the rules of the model. 
In Sec. V we give an explicit time dependent solution for $c(x,t)$ and in Sec. VI we give its scaling 
description to obtain various scaling exponents. Finally, in Sec. VII,
we discuss and summarize the work. 

\begin{figure}
\includegraphics[width=8.50cm,height=4.15cm,clip=true]{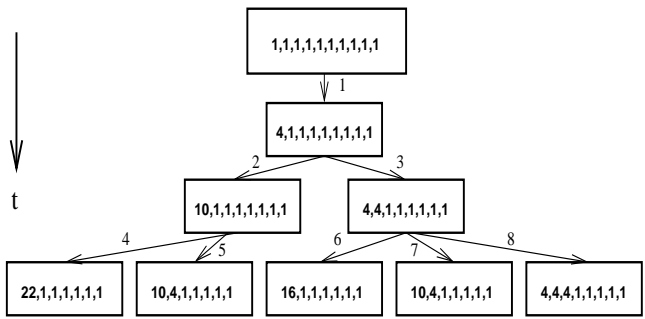}
\caption{Schematic representation of the model for $\alpha=1$.
}
\label{fig1}
 \end{figure}

\section{Model} 

We assume that initially the system consists of a large number of equal
sized chemically identical particles which are assumed to be immersed in the gas phase. 
These particles, while in Brownian motion, are 
continuously growing by heterogeneous condensation, leading to gas-to-particle conversion,
and merge irreversibly with other similarly growing particles upon encounter.
 To further specify our model, we assume
that the amount of net growth by condensation of a given particle between collisions, 
in the most generic case, is 
directly proportional to its own size. The algorithm for one time unit of the model can then be 
defined as follows: 
\begin{itemize}
\item[(i)] Two particles are picked randomly from the system which mimics 
random collision {\it vis-a-vis} Brownian motion in one dimension.
\item[(ii)] The sizes of the two particles are increased by a fraction $\alpha$ 
of their respective sizes. 
\item[(iii)] Their sizes are combined to form one particle. 
\item[(iv)] The steps $(i)-(iii)$ are repeated {\it ad infinitum}. 
\end{itemize}
While the model is rather simple to define, the results it offers are far from 
simple. In order to illustrate how the rules of the model work for 
mono-disperse initial conditions, we give a simple 
example in Fig. $1$ using an  evolutionary-tree-based approach. 
The state of the system at a given 
time is fully described by the numbers in the corresponding box. 
The evolution of these numbers is then described by a set 
of such boxes in successive times along the possible 
trajectory, e.g., $1\rightarrow 3\rightarrow 6\rightarrow$ and so 
on. The first salient feature of this model is that the sum of all 
the numbers $L$ in the successive boxes keep increasing continuously with 
$t$ revealing that the conservation of mass law is clearly violated. Secondly, 
the numbers in the different boxes 
represent the size or mass of the aggregate and hence a given box can be 
characterized by a distribution function $c(x,t)$ of particle of linear size $x$ at time $t$. 

\begin{figure}
\includegraphics[width=8.50cm,height=4.15cm,clip=true]{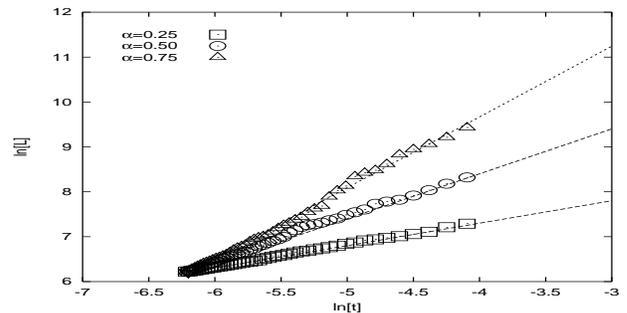}
\caption{$\ln[L]$ vs $\ln[t]$ (where $t=1/N$ since the number of particles present
ultimately determines how fast or slow the aggregation process should proceed)
are drawn using simulation data from one realization. The solid lines
represent theoretical predictions with gradient equal to $2\alpha$.
}
\label{fig2}
\end{figure}

\section{Numerical Simulation}

\begin{figure}
\includegraphics[width=8.50cm,height=4.15cm,clip=true]{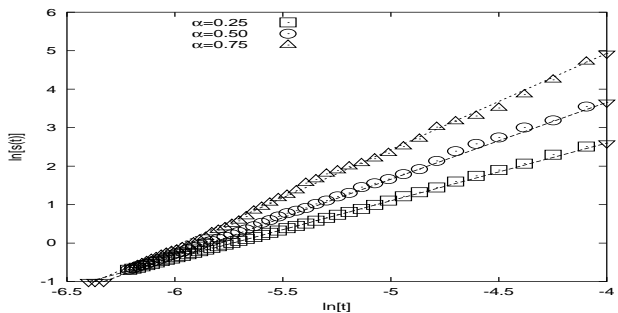}
\caption{$\ln [s]$ vs $\ln [t]$ are drawn using data collected from numerical simulation.
Data points (denoted by symbols) for a given curve represent one realization of our simulation.
 The solid lines
represent theoretical predictions with gradient equal to $1+2\alpha$.
}
\label{fig3}
\end{figure}

In order to extract a couple of basic features of the model, we have performed extensive 
numerical simulations based on the algorithm $(i)$-$(iv)$. Perhaps it is worthwhile to recall 
the work of Falk and Thomas, who in $1974$ obtained the 
molecular-size distribution by simulating the discrete version of Eq. (\ref{eq:s}) \cite{ref.falk}. 
The most crucial aspect of the stochastic simulation
is to find a way to define the time $t$. Note that the time necessary for a particle
to come into contact with another particle should depend on the number of particles 
present in the system. Indeed, ultimately it is the number density of particles present in the 
system which should determine how fast or how slowly the process should proceed, provided the 
collisions for aggregation of particles are independent of their size, and hence we define 
time $t=1/N$. In Fig \ref{fig2} we therefore have plotted
$\ln [L(t)]$ against $\ln [t]$ for different $\alpha$ and found a straight line with slope 
equals to $2\alpha$. This implies that
\begin{equation}
\label{eq:L}
L(t)\sim t^{2\alpha},
\end{equation}
which immediately confirms that, for systems describing the condensation-driven aggregation,
conservation of mass principle is always violated. 
In Fig. \ref{fig3} we also have plotted $\ln[s(t)]$ where $s(t)$ is the mean particle size 
defined as 
\begin{equation}
\label{eq:st}
s(t)=L(t)/N(t),
\end{equation}
against $\ln [t]$, and again found a straight line with slope equals to $1+2\alpha$ for all $\alpha>0$. We can thus write
the following growth law for the mean particle size:
\begin{equation}
\label{eq:s}
s(t)\sim t^{1+2\alpha}.
\end{equation}

\section{Analytical Model}

To solve the model analytically, we use the generalized Smoluchowski (GS) equation
\begin{eqnarray}
\label{eq:1}
\Big[{{\partial }\over{\partial t}} &+&  {{\partial}\over{\partial x}} v(x,t) \Big]c(x,t)
=-c(x,t)\int_0^\infty K(x,y)c(y,t)dy \nonumber \\ &+ & {{1}\over{2}}\int_0^x dy K(y,x-y)
c(y,t)c(x-y,t),
\end{eqnarray}
where $v(x,t)$ is the velocity with which particles grow by condensation. In the absence of 
the second term on the right hand side,
Eq. (\ref{eq:1}) reduces to the classical Smoluchowski (CS) equation as given in 
(Eq. \ref{eq:cs}) whose dynamics is governed by the conservation of mass law 
\cite{ref.smoluchowski}. The GS equation does not automatically describe our model
unless the expressions for the growth velocity $v(x,t)$, 
the collision time $\tau$, and the kernel $K(x,y)$ are suitably specified. 
To obtain a suitable expression for $v(x,t)$, 
it is worthwhile to recall the definition of the mean growth velocity which is defined as
\begin{equation}
\label{eq:velocity}
{\rm average \ growth \ velocity}={{{\rm net \ growth \ size} \ \Delta x}\over{{
\rm elapsed \ time}\ \Delta t}}.
\end{equation}
According to rule $(ii)$, the net growth of a particle of size $x$ between collisions is 
$\Delta x=\alpha x$.
On the other hand, a simple dimensional analysis of Eq. (\ref{eq:1}) reveals that 
the inverse of $\int_0^\infty K(x,y)c(y,t)dy$ is the collision time $\Delta t=\tau(x)$
during which the growth $\alpha x$ takes place. 
The mean growth velocity between collision therefore is 
\begin{equation}
\label{eq:2}
 v(x,t) = {{\alpha x}\over{\tau(x)}}=\alpha x\int_0^\infty dyK(x,y)c(y,t).
\end{equation} 
On the other hand, the rule $(i)$ says that a particle collides with any particle in the system,
irrespective of their size  with 
an equal {\it a priori} probability. We therefore should choose the collision kernel independent
of the size of the colliding particles i.e., we choose constant collision kernel or
\begin{equation}
\label{eq:3}
 K(x,y) = 2,
\end{equation}
for convenience.

To check our results from the numerical simulation, we now incorporate the $k^{th}$ 
moment defined as
\begin{equation}
\label{eq:4}
M_k(t)=\int_0^\infty x^kc(x,t)dx, \hspace{0.30cm} {\rm with} \hspace{0.25cm} k \geq 0,
\end{equation}
together with Eqs. (\ref{eq:2}) and (\ref{eq:3}) in Eq. (\ref{eq:1}) to write the rate equation for $M_k(t)$ 
in the closed form  
\begin{equation}
\label{eq:5}
 {{dM_k}\over{dt}} = 
\sum_{r=0}^k \left( \begin{array}{c} k \\ r \end{array}\right )
M_rM_{k-r} +  2(\alpha k - 1)M_0M_k,
\end{equation}
for integer $k$ value. We can readily solve it for the first moment $M_1(t)\equiv  L(t)$ to give
\begin{equation}
\label{eq:6}
 L(t) = (1 + N(0)t)^{2\alpha}.
\end{equation}
In the long time limit it grows following the same relation as in Eq. (\ref{eq:6}) which confirms a perfect matching with 
our numerical simulation.
On the other hand, solving Eq. (\ref{eq:5}) with $n=0$ we find that
\begin{equation}
\label{eq:number}
N(t)={{N(0)}\over{1+N(0)t}},
\end{equation}
and hence asymptotically the total number $N$ decays following the same power law 
\begin{equation}
N(t)\sim t^{-1},
\end{equation} 
with the same exponent as that of the CS equation. This is consistent with the assumption that the new particles are 
not nucleated. A similar temporal behavior has also been observed in other theories and 
experiments \cite{ref.friedlander,ref.husar}. One can use the two solutions for $L(t)$ and 
$N(t)$ in Eq. (\ref{eq:st}) to find the growth law for the mean particle size $s(t)$ 
 \begin{equation}
s(t)\sim (1+N(0)t)^{1+2\alpha}.
\end{equation}
Therefore, one can immediately find that in the long-time limit the mean particle grows algebraically 
following the same relation as in Eq. (\ref{eq:s}) which further confirms that the GS equation together with the
constant kernel and the growth velocity do describe the model in question.

\section{Exact Solution}

To solve GS equation exactly, we define the Laplace transformation $\psi(p,t)$ of $c(x,t)$ by
\begin{equation}
\label{eq:7}
 \psi(p,t) = \int_0^\infty dx e^{-px} c(x,t),
\end{equation}
and its inverse transform to obtain $c(x,t)$ by
\begin{equation}
\label{eq:8}
c(x,t) = {{1}\over{2\pi
i}}\int_{\gamma-i\infty}^{\gamma+i\infty}dpe^{px}\psi(p,t), 
\end{equation}
with $\Re(p)>\gamma$. Differentiating Eq. (\ref{eq:7}) with respect to $t$ 
and using Eq. (\ref{eq:1}),
we find that $\psi(p,t)$ obeys the following nonlinear partial differential equation  
\begin{equation}
\label{eq:9}
{{\partial \psi(p,t)}\over{\partial t}} +2N(t)\big[1-
\alpha p{{\partial}\over{\partial p}}\big]\psi(p,t) =
\psi^2(p,t),
\end{equation}
where, of course
\begin{equation}
\label{eq:10}
N(t)=\psi(0,t) =M_0(t).
\end{equation}
Consequently, we need to solve Eq. (\ref{eq:9}) subject to the
initial condition
\begin{equation}
\label{eq:11}
\psi(p,0) = \int_0^\infty dx e^{-px}c(x,0)\equiv f(p),
\end{equation}
We incorporate
the solution for $N(t)$ from Eq. (\ref{eq:number}) in Eq. (\ref{eq:9}) and then re-write 
it as follows
\begin{eqnarray}
\label{eq:12}
{{\partial \psi(p,t)}\over{\partial t}}&-& \Big ({{2\alpha N(0)p}\over{1+N(0)t}}\Big )
{{\partial\psi(p,t)}\over{\partial p}}\nonumber \\ &=& \psi^2(p,t)-{{2 N(0)}\over{1+N(0)t}}\psi(p,t).
\end{eqnarray}
To solve this equation we use the method of characteristic in which one usually writes
\begin{equation}
\label{eq:13}
{{d\psi}\over{ds}}={{\partial \psi}\over{\partial t}}{{\partial t}\over{\partial s}}+
{{\partial \psi}\over{\partial p}}{{\partial p}\over{\partial s}}.
\end{equation}
Comparing the above two equations we get 
\begin{eqnarray}
\label{eq:method1}
& & {{\partial t(s)}\over{\partial s}}= 1, \nonumber \\
& & {{\partial p(s)}\over{\partial s}}=-{{2\alpha N(0)}\over{1+N(0)t}}p(s). 
\end{eqnarray}
The quantity $\psi$ thus evolves following the ordinary non-linear equation
\begin{equation}
\label{eq:14}
{{d\psi}\over{ds}}=\psi^2-{{2 N(0)}\over{1+N(0)t(s)}}\psi.
\end{equation}
Solving Eqs. (\ref{eq:method1}) subject to initial data
\begin{eqnarray}
& & t(s=0)=0 \nonumber \\
& & p(s=0)=p_0 \nonumber \\
& & \psi(s=0)=f(p_0)
\end{eqnarray}
yields
\begin{eqnarray}
& & t=s \nonumber \\
& & p_0=p(1+N(0)s)^{2\alpha}.
\end{eqnarray}
We can transform Eq. (\ref{eq:14}) into a linear equation by setting  
\begin{equation}
\label{eq:chi}
\psi={{1}\over{\chi}},
\end{equation}
 to obtain
\begin{equation}
{{d\chi}\over{ds}}-{{2N(0)}\over{1+N(0)s}}\chi=-1.
\end{equation}
We solve it by using the integrating factor $I=(1+N(0)s)^{-2}$ and find that
 \begin{equation}
\chi(s,p_0)=(1+N(0)s)\Big ((1+N(0)s)\chi(0,p_0)-s\Big ),
\end{equation}
and hence using it back in Eq. (\ref{eq:chi}) we get
\begin{equation}
\label{eq:15}
\psi(p,t) = {{ f\big (p({1 + N(0)t})^{2\alpha}\big )}\over{ (1 + N(0)t)^2\Big (1 - 
{{  f  \big (p (1 + N(0)t)^{2\alpha}\big )t}\over{1+N(0)t}}\Big )}}.
\end{equation}
Note that, according to the definition of our model, we are interested in 
systems containing a large number of
chemically identical particles of unit size and hence without loss of generality 
we may choose the mono-disperse 
initial condition
\begin{equation}
c(x,0) = \delta (x - 1),
\end{equation}
which gives 
\begin{equation}
f(p)=e^{-p}, \hspace{0.30cm} {\rm and} \hspace{0.30cm} N(0)=1.
\end{equation}
To find $c(x,t)$ we substitute Eq. (\ref{eq:15}) in the definition of the inverse 
Laplace transform [(Eq. (\ref{eq:8})] and then a short calculation yields
\begin{equation}
\label{eq:16}
c(x,t) = {{ t^{(1+t)^{-2\alpha}x-1}\over{  (1+t
)^{2+2\alpha}(1+t)^{(1+t)^{-2\alpha}x-1}}}}, \hspace{0.35cm} \forall \ t>0.
\end{equation}
We can readily see that in the limit $\alpha \rightarrow 0$, we get the 
well-known solution of the CS equation \cite{ref.smoluchowski, ref.chandrasekhar}.

\section{Scaling Description}

Finding scaling or self-similar solutions is, more often than not, of utmost importance. 
These are essentially the 
solutions in the long-time limit where the distribution function $c(x,t)$ takes a simple 
universal form, in the sense that, it is independent of initial conditions. 
Most experimental systems do evolve to the point where such behavior is reached. Taking 
the limit $t\longrightarrow \infty$ and using the identity 
\begin{equation}
\lim_{n\rightarrow \infty} \Big(1+{{1}\over{n}}\Big)^{n}=e,
\end{equation}
in Eq. (\ref{eq:16}) we get
\begin{equation}
\label{eq:19}
c(x,t) \sim t^{-(2+2\alpha)}\exp[-{{x}\over{t^{1+2\alpha}}}].
\end{equation}
The structure of the solution given above is highly instructive as it satisfies 
\begin{equation}
\label{eq:affine}
c(b^{(1+2\alpha)}x,b t) =g(b)c(x,t),
\end{equation}
where
\begin{equation}
g(b)=b^{-(2+2\alpha)}.
\end{equation}
It implies that if we increase the units 
of measurement of $x$ by a factor $b^{(1+2\alpha)}$ and that of time by a 
factor of $b$, the numerical value of $c(x,t)$ is decreased by a factor of $g(b)$. 
The existence of scaling in the present case means that a plot of 
$c(x,t)/t^{-\beta}$ vs $x/t^z$ collapses into one graph for all initial
conditions. Mathematically, a solution of this kind
is called scale invariant and it implies that the system lacks a characteristic length scale.
A further testament to the fact that the solution given by Eq. (\ref{eq:19}) exhibits scaling is that it has 
exactly the same form
as in Eq. (\ref{eq:0}) and hence comparing the two we can extract the scaling exponents
\begin{equation}
\label{eq:mass}
\beta=2+2\alpha, \hspace{0.45cm} z=1+2\alpha,
\end{equation}
and the scaling function
\begin{equation}
\phi(\xi)=e^{-\xi}.
\end{equation}

We can now substitute Eq. (\ref{eq:19}) in Eq. (\ref{eq:4}) to obtain  
the asymptotic solution for the $k^{th}$ moment
\begin{equation}
\label{eq:20}
M_k(t) \sim t^{-\gamma(k)},
\end{equation}
where,
\begin{equation}
\gamma(k)=1-(2\alpha + 1)k.
\end{equation}
We thus see that the exponent $\gamma(k)$ is linear in $k$ which means that there 
exists a constant gap between 
exponents of consecutive $k$, and hence we can define the mean cluster size 
\begin{equation}
s(t)={{M_k(t)}\over{M_{k-1}(t)}} \hspace{0.35cm} {\rm for \ integer} \ 
k\geq 1,
\end{equation}
The mean particle size $s(t)$ therefore grows algebraically 
with an exponent equal to 
\begin{equation}
\gamma(k)-\gamma(k-1)=1+2\alpha,
\end{equation}
 which is exactly the same as in Eq. (\ref{eq:s})
obtained by numerical simulation. 
Using Eq. (\ref{eq:s})  in Eq. (\ref{eq:0}), we can therefore write yet another widely used 
form of the scaling anzatz,
\begin{equation}
c(x,t)\sim s(t)^{-\theta}\phi\big (x/s(t)\big ),
\end{equation}
with the mass exponent $\theta =\beta/z$, and hence
according to Eq. (\ref{eq:mass}) we get
\begin{equation}
\label{eq:massexponent}
\theta=\frac{2+2\alpha}{1+2\alpha}.
\end{equation}
The expression for the exponent $\gamma(q)$ reveals that the moment $M_q(t)$ becomes time independent 
if 
\begin{equation}
q={{1}\over{1+2\alpha}} \hspace{0.3cm} \forall \ \alpha>0,
\end{equation}
since $\gamma(q)=0$. Incorporating the value of $q$ in Eq. (\ref{eq:massexponent})
we immediately find that
\begin{equation}
\label{eq:23}
\theta=1+q,
\end{equation}
which is always less than $2$ for all $\alpha>0$. We can recover the classical value $\theta=2$ of aggregation
without condensation by setting $\alpha=0$.

\section{Discussion and Summary}

One may solve the GS equation for other growth velocities following the same 
method. Culille and Sire, in fact, solved
the GS equation for $v=1$ and $v=x$ exactly and found  
\begin{equation}
\label{eq:sire}
c(x,t)\sim {{2}\over{t^2\ln t}}e^{-{{x}\over{t\ln t}}}; \hspace{0.35cm} c(x,t)\sim {{4}\over{t^2e^t}}
e^{-{{2x}\over{te^t}}},
\end{equation}
respectively for constant kernel $K(x,y)=1$ \cite{ref.sire}. 
As we have the exact solutions for three different
growth velocities, we find it worthwhile to compare them and 
see their dissimilarities. First of all, note that one cannot choose $v=1$ and $v=x$ and 
claim them to describe the growth velocities between collisions since the elapsed time 
$\Delta t$ during which the net growth occurs
has not been chosen to be equal to the mean collision time $\tau=1/\int_0^\infty K(x,y)c(y,t)dy$. 
Second,  neither of the 
two solutions given in Eq. (\ref{eq:sire}) can be expressed in the form of 
Eq. (\ref{eq:affine}) or in the form of Eq. (\ref{eq:0}) and hence they violate scaling. 
Third, one finds that the 
$k$th moment of both solutions ($v=1$ and $v=x$) no longer exhibits power-law against time $t$
 which is a further testament to the violation of scaling since power-law relations for various
moments of the distribution function $c(x,t)$ is the {\it hall mark} for the 
existence of scaling. Finally, when we choose $v=0$ and $v=\alpha x/\tau$, 
the dynamics of the systems are governed by some conservation laws and the scaling exponents are in fact fixed by 
these laws. However, there is no such conservation law in the case of $v=1$ or $v=x$ and since the scaling
is violated and hence exponents are nonexistent. 

In this paper, we have presented an exactly solvable analytical model to study the kinetics 
of irreversible aggregation of particles growing by heterogeneous condensation with velocity 
$v=\alpha x/\tau$. As a result of the additional growth by condensation, we found an 
algebraic growth law for the mean particle size $s(t)\sim t^{1+2\alpha}$
instead of a linear growth $s(t)\sim t$ in the absence of growth by condensation.
The size spectra of the aggregates formed by the condensation-driven 
aggregation is shown to exhibit universal scaling $c(x,t)\sim t^{-\theta z}\phi(x/t^z)$ with mass
exponent $\theta=1+q$ which is always less than its classical counterpart $\theta=2$,
 and kinetic exponent $z=1+2\alpha$.
We have found the exact and explicit solution for the scaling function 
$\phi(\xi)\sim e^{-\xi}$. Interestingly,
one obtain the same scaling function $e^{-\xi}$ for aggregation without condensation. 
The difference
appears only in the $z$ values of the scaling argument $\xi=x/t^z$ since the two systems have 
different $z$ values. In this sence, the form of the scaling function is universal in nature.
We have shown that the transition to such scaling is 
accompanied by the emergence of a non-trivial conservation law, that is, $M_q(t)\sim$ constant.
We believe that the present work will attract a renewed interest in the subject. The ideas developed
in this paper could be taken further by investigating the same model for
generalized homogeneous aggregation kernel $K(x,y) = (xy)^{\lambda/2}$ which we intend to do in our future endeavor.


\begin{thebibliography}{99}

\bibitem{ref.friedlander} S. K. Friedlander, {\it Smoke, Dust and Haze}, (New York, Wiley, 1977).
\bibitem{ref.thorn} M. Thorn and M. Seesselberg, Phys. Rev. Lett. {\bf 72} 3622 (1994); M. L. Broide and R. J. Cohen, Phys. Rev. Lett. {\bf 64} 2026 (1990).
\bibitem{ref.melle} S. Melle, M. A. Rubio, and G. G. Fuller, Phys. Rev. Lett. {\bf 87} 115501 (2001).
\bibitem{ref.polymerization} E. Ben-Naim and P. L. Krapivsky, J. Phys. Cond. Mat. {\bf 17} S4249 (2005).
\bibitem{ref.antigen}  D. Johnstone and G. Benedek, {\it Kinetics of Aggregation and Gelation}, edited by F. 
Family and D. P. Landau (North-Holland, Amsterdam, 1984).
\bibitem{ref.galaxy} J. Silk and S. D. White, Astrophys. J. {\bf 223} L59 (1978).
\bibitem{ref.smoluchowski} M. von Smoluchowski, Z. Phys. {\bf 92}, 129 (1917).
\bibitem{ref.chandrasekhar} S. Chandrasekhar, Rev. Mod. Phys. {\bf 15}, 1, (1943).
\bibitem{ref.ziff} R. M. Ziff, E. M. Hendriks and, M. H. Ernst, Phys. Rev. Lett. {\bf 49}, 593 (1982); P. G. J. 
van Dongen and M. H. Ernst, Phys. Rev. Lett. {\bf 54}, 1396 (1985).
\bibitem{ref.lushnikov} A. A. Lushnikov,  Phys. Rev. Lett. {\bf 93} 198302 (2005).
\bibitem{ref.vicsek} T. Vicsek, {\it Fractal Growth Phenomena}, 2nd ed. (World Scientific, Singapore, 1992).
\bibitem{ref.scaleinvariance} {\it Scale Invariance, Interfaces, and Non-Equilibrium Dynamics}, eidted by A. McKane, M. Droz, J. Vannimenus and D. Wolf, Nato ASI Series B {\bf 344} (1995).
\bibitem{ref.scaling} Z. Cheng and S. Redner, Phys. Rev. Lett. {\bf 60} 2450 (1988).
\bibitem{ref.droplet} P. L. Krapivsky and S. Redner, Phys. Rev. E {\bf 54}, 3553 (1996). 
A. A. Lushnikov and M. Kulmala, Phys. Rev. E {\bf 63} 061109 (2001).
\bibitem{ref.sire} S. Cuille and C. Sire, Europhys. Lett. {\bf 40}, 239 (1997); {\it ibid}, Phys. Rev. E {\bf 57}
881 (1998).
\bibitem{ref.husar} P. Meakin and F. Family, J. Phys. A: Math. Gen., {\bf 22} L225 (1989). 
\bibitem{ref.dust} P. Tullet, Phys. Educ. {\bf 34} 140 (1999). 
\bibitem{ref.falk} M. Falk and R. E. Thomas, Can. J. Chem. {\bf 52} 3285 (1974).
\end{thebibliography}
\end{document}